\documentclass[12pt]{article}
\usepackage{graphicx}
\usepackage{amsmath}
\date{}
\begin{document}

\title{Long's Equation in Terrain Following Coordinates}

\author{Mayer Humi\\
Department of Mathematical Sciences\\
Worcester Polytechnic Institute\\
100 Institute Road\\
Worcester, MA  01609}

\maketitle
\thispagestyle{empty}

\begin{abstract}

Long's equation describes two dimensional stratified atmospheric
flow over terrain which is represented by the geometry of the domain. 
The solutions of this equation over simple topography were investigated 
analytically and numerically by many authors. In this paper we derive 
a new terrain following formulation of this equation which incorporates  
the terrain as part of the differential equation rather than the 
geometry of the domain. This leads to new analytic insights about 
the solutions of this equation and enable us to compute 
steady state gravity wave patterns over complex topography.
\end{abstract}

\vfill

 
\newpage

\setcounter{equation}{0}
\section{Introduction}

Long's equation [1,2,3,4] models the flow of stratified 
incompressible fluid in two dimensions 
over terrain. When the base state of the flow (that is the unperturbed flow
field far upstream) is without shear the numerical solutions (in the 
form of steady lee waves) of this equation over simple topography 
(i.e. one hill) were studied by many authors [5-13].  
The most common approximation in these studies was to set 
Brunt-V\"{a}is\"{a}l\"{a} frequency to a constant 
or a step function over the computational domain.
Moreover the values of two physical parameters which appear in this
equation were set to zero. (These parameters control the 
stratification and dispersive effects of the atmosphere - see $Sec\,2$.)
In this (singular) limit the nonlinear 
terms and one of the leading second order derivatives in the equation drop 
out and the equation reduces to that of a linear harmonic oscillator over 
two dimensional domain. Careful studies [8] showed that these approximations
set strong limitations on the validity of the derived solutions [9].

Long's equation also provides the theoretical framework for the analysis
of experimental data  [14,15,29] under the assumption of shearless base 
flow. (An assumption which, in general, is not supported by the data). 
An extensive list of references appears in [16,17,18].

An analytic approach to the study of the solutions of this nonlinear 
equation was initiated recently by the current author [19,20,21]. We showed 
that for a base flow without shear and under rather mild restrictions 
the nonlinear terms in the equation can be simplified. We also identified the
"slow variable" that controls the nonlinear oscillations in this equation.
Using phase averaging approximation we derived for self similar solutions
of this equation a formula for the attenuation of the stream function 
perturbation with height. This result is generically related to the 
presence of the nonlinear terms in Long's equation. The impact that
shear has on the generation and amplitude of gravity waves was 
investigated by us in [20]. A new representation of this equation in terms
of the atmospheric density was derived in [21].

One of the weak aspects of Long's equation is related to the fact that
the terrain is represented by the shape of the domain and the boundary
conditions. As a result the impact of different terrains on the solution
of this equation can only be studied numerically. Furthermore  
discretization errors which occur in the representation of the terrain
render it impractical to consider complex terrain. In part these errors
are due to the scale of the terrain relative to the computational domain.
Accordingly only simple topographies which were represented by one hill were
considered in the literature. Furthermore even for these simple topographies
only approximate boundary conditions were applied at the terrain.
(See discussion in $Sec\; 2$).

With this motivation it is our objective in this paper to derive 
a terrain following formulation of Long's equation in which the terrain
is incorporated as part of the coefficients of the differential equation,
and the computational domain is always a rectangle. This new 
representation makes it possible to derive new analytic insights about
the solution of this equation in some limiting cases. It will make it
easier also to study how the solution varies as a function of the terrain
and other parameters that appear in the equation.
 
The plan of the paper is as follows: Sec. 2 presents a short 
review of Long's equation and some aspects of its solutions.
In Sec. 3 we derive the new formulation of this equation. 
Sec 4 considers some analytic and geophysical aspects of this new 
formulation while Sec 5 compares its numerical solution over three different 
terrains. These simulations are motivated by recent experiments in the 
Alps region to educe properties of gravity waves from experimental data [14].
We end up in Sec 6. with a summary and conclusions. 

\setcounter{equation}{0}
\section{Long's Equation - A Short Overview}

In two dimensions $(x,z)$ the flow of a steady inviscid and incompressible
stratified fluid is modeled by the
following equations:
\begin{equation}
\label{2.1}
u_x + w_z = 0
\end{equation}
\begin{equation}
\label{2.2}
u\rho_x + w\rho_z = 0
\end{equation}
\begin{equation}
\label{2.3}
\rho(uu_x+wu_z) = -p_x
\end{equation}
\begin{equation}
\label{2.4}
\rho(uw_x+ww_z) = -p_z-\rho g
\end{equation}
where subscripts indicate differentiation with respect to the indicated
variable, ${\bf u}=(u,w)$ is the fluid velocity, $\rho$ is its density $p$ 
is the pressure and $g$ is the acceleration of gravity.

We can non-dimensionalize these equations by introducing
\begin{eqnarray}
\label{2.5}
\bar{x} &=& \frac{x}{L},\;\;\bar{z} = \frac{N_0}{U_0} z,\;\;\bar{u} =
\frac{u}{U_0},\;\;
\bar{w} = \frac{LN_0}{U^2_0} w \nonumber \\
\bar{\rho} &=& \frac{\rho}{\bar\rho_0},\;\;\bar{p} = \frac{N_0}{gU_0
{\bar\rho_0}}p
\end{eqnarray}
where $L$ represents a characteristic horizontal length, and $U_0,\bar{\rho_0}$
represent respectively the free stream velocity and averaged base density 
(i.e. here $\bar\rho_0$ is a constant). $N_0^2$ is 
an averaged value of the Brunt-V\"{a}is\"{a}l\"{a} frequency 
\begin{equation}
\label{2.6}
N^{2} = -\frac{g}{\rho_0}\;\frac{d\rho_0}{dz}
\end{equation}
where $\rho_0=\rho_0(z)$ is the base density.

In these new variables eqs (\ref{2.1})-(\ref{2.4}) take the following form 
(for brevity we drop the bars)
\begin{equation}
\label{2.7}
u_x + w_z = 0
\end{equation}
\begin{equation}
\label{2.8}
u\rho_x + w\rho_z = 0
\end{equation}
\begin{equation}
\label{2.9}
\beta\rho(uu_x + wu_z) = -p_x
\end{equation}
\begin{equation}
\label{2.10}
\beta\rho(uw_x + ww_z) = -\mu^{-2}(p_z + \rho)
\end{equation}
where 
\begin{equation}
\label{2.11}
\beta = \frac{N_0U_0}{g}
\end{equation}
\begin{equation}
\label{2.12}
\mu = \frac{U_0}{N_0L}\;.
\end{equation}
$\beta$ is the Boussinesq parameter [13] (this name has nothing to do with
the "Boussinesq approximation") which controls 
stratification effects (assuming $U_0 \neq 0$) and $\mu$ is the long wave 
parameter which controls dispersive effects (or the deviation from the 
hydrostatic approximation). In the limit $\mu=0$ the hydrostatic 
approximation is fully satisfied, [10,11].

In view of eq. (\ref{2.7}) we can introduce a stream function $\Psi$ so that
\begin{equation}
\label{2.13}
u = \Psi_z,\;\;w = -\Psi_x\;.
\end{equation}

Using this stream function we can rewrite eq. (\ref{2.8}) as
\begin{equation}
\label{2.14}
J\{\rho,\Psi \}=0
\end{equation}
where for any two (smooth) functions $f,g$
\begin{equation}
\label{2.15}
J\{f,g\}=\frac{\partial f}{\partial x}\frac{\partial g}{\partial z} -
        \frac{\partial f}{\partial z}\frac{\partial g}{\partial x}
\end{equation}
Eq. (\ref{2.14}) implies that the functions $\rho,\Psi$ are dependent
on each other and we can express each of them in terms of the other.
Thus we can write $\Psi$ as $\Psi(\rho)$ (or $\rho$ as $\rho(\Psi)$ [21]).

After a long algebra one can derive the following equation 
for $\Psi$ [22,1,13]
\begin{equation}
\label{2.21}
\Psi_{zz} + \mu^2\Psi_{xx}-N^2(\Psi)\left[z+\frac{\beta}{2}(\Psi^2_z +
  \mu^2\Psi^2_x)\right]=S(\Psi)
\end{equation}
where
\begin{equation}
\label{2.22}
N^2(\Psi) = - \frac{\rho_\Psi}{\beta\rho}
\end{equation}
is the nondimensional Brunt-V\"{a}is\"{a}l\"{a} frequency. We observe that
in this definition $N^2$ is a function of $\Psi$. (As a result it can be an
additional source of nonlinearity in eq. (\ref{2.21}))  
This is in contrast to the previous definition of this quantity in 
eq. (\ref{2.6}) which depends only on the base state. In the following
we assume without loss of generality that the direction of base flow is
from left to right along the x-axis. Furthermore 
we assume it to be a function of $z$ only.

$S(\Psi)$ is some unknown function which is determined from the base 
flow.  To carry out this determination of $S$ we consider eq. 
(\ref{2.21}) as $x \rightarrow -\infty$ and express 
the left hand side of this equation 
in terms of $\Psi$ only. (Assuming that disturbances do not propagate far 
upstream [16,18]). Eq. (\ref{2.21}) is referred to as Long's equation. 

For example if we let
\begin{equation}
\label{2.23}
\displaystyle\lim_{x \rightarrow -\infty}\Psi(x,z) = z
\end{equation}
i.e consider a shearless base flow with 
$\displaystyle\lim_{x \rightarrow -\infty} u(x,z) =1$ then 
\begin{equation}
\label{2.24}
S(\Psi) = -N^2(\Psi)\left(\Psi + \frac{\beta}{2}\right)
\end{equation}
and eq. (\ref{2.21}) becomes:
\begin{equation}
\label{2.25}
\Psi_{zz} +
\mu^2\Psi_{xx}-N^2(\Psi)[z-\Psi+\frac{\beta}{2}(\Psi^2_z+\mu^2\Psi^2_x
  -1)] = 0.
\end{equation}
It is evident from this derivation that different profiles for the base 
flow as $x \rightarrow -\infty$ will lead to different forms of $S(\Psi) [20]$. 

For a general base flow in an unbounded domain over topography with shape 
$f(x)$ and maximum height $H_0$ the following boundary conditions are imposed 
on $\Psi$ 
\begin{equation}
\label{2.26}
\displaystyle\lim_{x \rightarrow -\infty}\Psi(x,z) = \Psi_0(z)
\end{equation}
\begin{equation}
\label{2.27}
\Psi(x,\tau f(x)) = \mbox{constant},\;\;\;\tau =
\frac{H_0N_0}{U_0}
\end{equation}
where the constant in eq.(\ref{2.27}) is (usually) set to zero. 
As to the boundary condition at $x \rightarrow \infty$ it is appropriate to set 
$$
\displaystyle\lim_{x \rightarrow \infty} \Psi (x,z) = \Psi_0(z)
$$
(in spite of the fact that Long's equation contains no dissipation terms). 
However over finite computational domain only radiation 
boundary conditions can be imposed in this limit. Similarly as 
$z \rightarrow \infty$ 
it is customary to impose (following [7]) radiation boundary conditions.
(The imposition of these boundary conditions is discussed in detail 
in $Sec\; 4.1$). 

For the perturbation from the shearless base flow 
\begin{equation}
\label{2.28}
\eta = \Psi - z
\end{equation}
eq. (\ref{2.25}) becomes
\begin{equation}
\label{2.29}
\eta_{zz} -
\alpha^2\eta^2_z+\mu^2(\eta_{xx}-\alpha^2\eta^2_x)-N^2(\eta)(\beta\eta_z-\eta)
= 0
\end{equation}
where
\begin{equation}
\label{2.30}
\alpha^2 = \frac{N^2 (\Psi)\beta}{2}\;.
\end{equation}

We observe that when $|\tau| \ll 1$ the boundary condition (\ref{2.27}) 
can be approximated by 
\begin{equation}
\label{2.31}
\eta(x,0)=-\tau f(x).
\end{equation}

When $N$ is constant eq. (\ref{2.29}) is invariant with respect to 
translations in $x,z$
and hence admits self-similar solutions of the form $\eta=f(kx+mz)$\, [19].
These solutions are interpreted as gravity waves that are generated
by the flow over the topography.

From a numerical point of view it is a common practice [7,8,13] 
to solve eq. (\ref{2.29}) in the limit $\beta=0$ and $\mu=0$ with 
constant $N$ over the domain. However observe that the definition of $N$ in
Long equation is given by (\ref{2.22}) and it depends on $\Psi$. 
In some other numerical simulations the computational domain is divided 
into subdomains where $N$ is constant in each subdomain but this led to 
numerical instabilities at the interface between these subdomains.

In these limits Eq. (\ref{2.29}) reduces then to a linear equation    
\begin{equation}
\label{2.32}
\eta_{zz} + N^2\eta = 0\;.
\end{equation}
We observe that the limit $\beta=0$ can be obtained either by letting 
$U_0 \rightarrow 0$ or $N_0 \rightarrow 0$. In the following we assume 
that this limit is obtained as $U_0 \rightarrow 0$ (so that stratification 
persists in this limit and the leading term in $N_0$ is not zero).

Eq. (\ref{2.32}) is  a singular limit of Long's 
equation as one of the leading second order derivatives drops when 
$\mu=0$ and the nonlinear terms drops out when $\beta=0$ and $N$ is constant. 
This approximation 
and its limitations were considered numerically and analytically [6,7,19,20] 
and was found to be justified only under strong restrictions even under 
the assumption that the base flow is shearless. Nevertheless it is used 
routinely in the actual analysis of atmospheric data [14,15,16].

The general solution of eq. (\ref{2.32}) is
\begin{equation}
\label{2.33}
\eta(x,z) = q(x)\cos(Nz)+p(x)\sin(Nz)
\end{equation}
where the functions $p(x),q(x$) have to be determined so that the the boundary
conditions derived from eq. (\ref{2.27}),(\ref{2.31}) and the radiation
boundary conditions are satisfied. These lead in general
to an integral equation for $p(x)$ and $q(x)$ and it easy to show [13] that
$p(x)=H[q(x)]$ where $H[q(x)]$ is the Hilbert transform of $q(x)$. The 
boundary condition on the terrain becomes;
\begin{equation}
\label{2.34}
q(x)\cos(\tau Nf(x)) + H[q(x)]\sin(\tau Nf(x)) = -\tau
f(x)\;.
\end{equation}
This integral equation has to be solved numerically [6,7,13].

\setcounter{equation}{0}
\section{Terrain Following Formulation.} 

To derive a terrain following formulation of Long's equation which incorporates
the terrain in the coefficients of the differential equation (rather than 
the shape of the domain) we introduce Gal-Chen transformation. If the 
height of the (bottom) terrain is described by a sufficiently smooth function 
$z=h(x)$ and the height of the computational flow region is finite, i.e. 
$h(x) \leq z \leq H$, where  $H$  is a constant, then this transformation 
is given by
\begin{equation}
\label{3.1}
\bar{x} = x,\;\;\;\bar{z} = H\displaystyle\frac{z-h(x)}{H-h(x)}.
\end{equation}
Under this transformation we have
\begin{equation}
\label{3.2}
\displaystyle\frac{\partial}{\partial x} = \frac{\partial}{\partial \bar{x}} 
+ G^{12}\frac{\partial}{\partial\bar{z}},\;\;\displaystyle\frac{\partial}
{\partial z} = \frac{1}{\sqrt{G}}\frac{\partial}{\partial \bar{z}}
\end{equation}
where
\begin{equation}
\label{3.3}
\displaystyle\frac{1}{\sqrt{G}} = \frac{H}{H-h(x)},\;\;\;G^{12} = 
\displaystyle\frac{1}{\sqrt{G}}\left(\frac{\bar{z}}{H} -1\right)h^\prime(x).
\end{equation}
Furthermore the expression of the Laplace operator becomes
\begin{equation}
\label{3.4}
\bar{\nabla}^2 = \displaystyle\frac{\partial^2}{\partial\bar{x}^2} + 
\left[\frac{1}{G} + (G^{12})^2\right]\displaystyle\frac{\partial^2}
{\partial\bar{z}^2} + 2G^{12}\displaystyle\frac{\partial^2}{\partial\bar{x}
\partial\bar{z}} + \left[\displaystyle\frac{\partial G^{12}}{\partial\bar{x}} 
+ G^{12} \displaystyle\frac{\partial G^{12}}{\partial\bar{z}} \right]
\displaystyle\frac{\partial}{\partial\bar{z}}.
\end{equation}

Under this transformation the continuity equation (\ref{2.7}) becomes
\begin{equation}
\label{3.5}
\displaystyle\frac{\partial u}{\partial \bar{x}} + G^{12} 
\displaystyle\frac{\partial u}{\partial \bar{z}} + \frac{1}{\sqrt{G}}
\frac{\partial w}{\partial \bar{z}} = 0.
\end{equation}
However if we introduce
\begin{equation}
\label{3.6}
v = \displaystyle\frac{1}{\sqrt{G}}(w + \sqrt{G} G^{12} u)
\end{equation}
then it is a simple algebra to show that eq. (\ref{3.5}) can be rewritten as
\begin{equation}
\label{3.7}
\displaystyle\frac{\partial}{\partial \bar{x}}(\sqrt{G}u) + 
\displaystyle\frac{\partial}{\partial \bar{z}}(\sqrt{G} v) = 0.
\end{equation}
From this equation we see that we can introduce a "terrain following
stream function"  $\psi$  so that
\begin{equation}
\label{3.8}
\bar{u} = \sqrt{G} u = \displaystyle\frac{\partial \psi}{\partial \bar{z}},
\;\;\;\; {\bar v} = \sqrt{G} v = - \displaystyle\frac{\partial \psi}
{\partial \bar{x}}.
\end{equation}
Multiplying eq. (\ref{2.8}) by $\sqrt{G}$ we can rewrite this equation
in the following form:
\begin{equation}
\label{3.9}
\bar{u} \displaystyle\frac{\partial \rho}{\partial \bar{x}} +
{\bar v} \displaystyle\frac{\partial \rho}{\partial \bar{z}} =0 .
\end{equation}
Using eq. (\ref{3.8}) this can be rewritten as
\begin{equation}
\label{3.10}
\bar{J}\{\rho,\psi\} =0  
\end{equation}
where $\bar{J}$ is defined as in eq. (\ref{2.15}) but with differentiations
with respect to $(\bar{x},\bar{z})$. Equation (\ref{3.10}) implies that
$\rho(\bar{x},\bar{z})=\rho(\psi(\bar{x},\bar{z}))$  (and vice versa).

To eliminate the pressure term from eqs. (\ref{2.9}) and (\ref{2.10})
we differentiate (\ref{2.9}) by $z$ and apply the operator $\mu^2 
\frac{\partial}{\partial x}$ to (\ref{2.10}) and subtract. We obtain
\begin{equation}
\label{3.11}
\beta \mu^2\rho_x(uw_x+ww_z) - \beta \rho_z(uu_x+wu_z) +
\beta \mu^2\rho(uw_x+ww_z)_x - \beta \rho(uu_x +wu_z)_z = -\rho_x .
\end{equation}
Using eq. (\ref{2.8}) the first two terms in this equation can be written as
\begin{eqnarray}
\label{3.12}
&&\beta \mu^2\rho_x(uw_x+ww_z) - \beta \rho_z(uu_x+wu_z) = \beta\left[\mu^2(-\rho_zww_x+\rho_xww_z)-\rho_zuu_x+\rho_xuu_z\right] \\ \nonumber
&&=\frac{\beta}{2}\left[\rho_x(u^2+\mu^2w^2)_z -\rho_z(u^2+\mu^2w^2)_x\right]
=\frac{\beta}{2\sqrt{G}}\left[\rho_{\bar{x}}(u^2+\mu^2w^2)_{\bar{z}}-
\rho_{\bar{z}}(u^2+\mu^2w^2)_{\bar{x}}\right] = \\ \nonumber
&&\frac{\beta}{2\sqrt{G}}\bar{J}\{\rho,u^2+\mu^2w^2\}.
\end{eqnarray}
Using eqs (\ref{3.6}) and (\ref{3.8}) to re-express $u^2+\mu^2w^2$
we have 
\begin{equation}
\label{3.13}
\frac{\beta}{2\sqrt{G}}\bar{J}\{\rho,u^2+\mu^2w^2\}=
\frac{\beta}{2\sqrt{G}}\rho_{\psi}\bar{J}\left\{\psi,\mu^2(\psi_{\bar{x}})^2+
2\mu^2G^{12}\psi_{\bar{x}}\psi_{\bar{z}} +\left[\frac{1}{G}+\mu^2(G^{12})^2
\right](\psi_{\bar{z}})^2 \right\}
\end{equation}

The third and the fourth terms in eq. (\ref{3.11}) can be rewritten using
(\ref{2.7}) as
\begin{equation}
\label{3.14}
\beta \mu^2\rho(uw_x+ww_z)_x - \beta \rho(uu_x +wu_z)_z=
\beta\rho\left[u(\mu^2w_x-u_z)+v(\mu^2w_x-u_z)\right]
=-\frac{\beta\rho}{\sqrt{G}}\bar{J}\{\psi,\chi\}
\end{equation}
where $\chi=\mu^2w_x-u_z$ is the vorticity.
Expressing $\chi$ in terms of $\psi$ we have
\begin{equation}
\label{3.15}
\chi =-\bar{\nabla}^2_{\mu}\psi
\end{equation}
where
\begin{equation}
\label{3.16}
\bar{\nabla}^2_{\mu} =
\mu^2\left\{ \displaystyle\frac{\partial^2}{\partial\bar{x}^2} + 2G^{12}\displaystyle\frac{\partial^2}{\partial\bar{x} \partial\bar{z}} + \left[\displaystyle\frac{\partial G^{12}}{\partial\bar{x}} + G^{12} \displaystyle\frac{\partial G^{12}}{\partial\bar{z}} \right] \displaystyle\frac{\partial}{\partial\bar{z}}\right\}+
\left[\frac{1}{G} + \mu^2(G^{12})^2\right]\displaystyle\frac{\partial^2}
{\partial\bar{z}^2} 
\end{equation}
is the "terrain following Laplace operator".

Finally for the right hand side of eq. (\ref{3.11}) we have 
\begin{equation}
\label{3.17}
-\rho_x=-\frac{1}{\sqrt{G}}\bar{J}\{\rho,g\}=-\frac{\rho_{\psi}}
{\sqrt{G}}\bar{J}\{\psi,g\}
\end{equation}
where 
$$
g(\bar{x},\bar{z})=\bar{z}+h(\bar{x})\left(1-\frac{\bar{z}}{H}\right)
$$

Combining all the results contained in eqs. (\ref{3.12})-(\ref{3.17}) we
can re-express eq. (\ref{3.11}) in the following form:
\begin{equation}
\label{3.18}
\bar{J}\left\{\psi,\displaystyle\bar{\nabla}^2_{\mu}\psi - \frac{N^2(\psi)\beta}{2}\left[\mu^2(\psi_{\bar{x}})^2+ 2\mu^2G^{12}\psi_{\bar{x}}\psi_{\bar{z}} + \left(\frac{1}{G}+\mu^2(G^{12})^2 \right)(\psi_{\bar{z}})^2 \right] -N^2(\psi)g(\bar{x},\bar{z})\right\} =0
\end{equation}
where $N^2(\psi)$ is defined as in eq. (\ref{2.22}). Hence it follows that,
\begin{equation}
\label{3.19}
\displaystyle\bar{\nabla}^2_{\mu}\psi - \frac{N^2(\psi)\beta}{2}\left[\mu^2(\psi_{\bar{x}})^2+ 2\mu^2G^{12}\psi_{\bar{x}}\psi_{\bar{z}} + \left(\frac{1}{G}+\mu^2(G^{12})^2 \right)(\psi_{\bar{z}})^2 \right] -N^2(\psi)g(\bar{x},\bar{z}) = S(\psi). 
\end{equation}
This is the terrain following form of Long's equation.
At this juncture it might be asked why one can not "save" this derivation
and apply the terrain following transformation (\ref{3.1}) directly to 
(\ref{2.21}). Doing so will yield an extremely complicated equation. 
This has been avoided in our derivation by the use of the "terrain 
following stream function" in (\ref{3.8}). 

To determine the function $S(\psi)$ in eq. (\ref{3.19}) we assume that
$$
\displaystyle\lim_{\bar{x} \rightarrow -\infty} h(\bar{x}) =0
$$
and that (as an example) $\psi$ satisfies
$$
\displaystyle\lim_{x \rightarrow -\infty} \psi(\bar{x},\bar{z}) = \bar{z}.
$$
It follows then that 
\begin{equation}
\label{3.20}
S(\psi) = -N^2(\psi)\left(\psi + \frac{\beta}{2}\right)
\end{equation}
and Long's equation becomes:
\begin{eqnarray}
\label{3.21}
&&\displaystyle\bar{\nabla}^2_{\mu}\psi - \frac{N^2(\psi)\beta}{2}\left[\mu^2(\psi_{\bar{x}})^2+ 2\mu^2G^{12}\psi_{\bar{x}}\psi_{\bar{z}} + \left(\frac{1}{G}+\mu^2(G^{12})^2 \right)(\psi_{\bar{z}})^2 \right]  \\ \nonumber
&&-N^2(\psi)\left[g(\bar{x},\bar{z})-\psi-\frac{\beta}{2}\right] = 0. 
\end{eqnarray}

In this representation the flow domain is a rectangle 
$[a,b]\times [0,H]$ or an infinite stripe $[-\infty,\infty] \times [0,H]$.
The boundary condition at the bottom topography is 
$$
{\bf u}\cdot {\bf n} =0
$$
where ${\bf n}$ is the normal to the topography which is described by
the curve $h(x)$. Hence this normal is given by ${\bf n} = (-h^{\prime}(x),1)$.
Using (\ref{3.6}),(\ref{3.8}) this leads to the boundary condition
\begin{equation}
\label{3.22}
\psi(\bar{x},0) = \mbox{constant}
\end{equation}
and this constant can be chosen to be zero. The other boundary condition 
that has to be imposed on $\psi$ is a radiation boundary condition at 
$\bar{z}= H$ (which implies that the outgoing wave is not reflected by the 
boundary). 

To obtain an equation for the perturbation from the base state we set
\begin{equation}
\label{3.23}
\psi(\bar{x},\bar{z}) = \bar{z} + \eta(\bar{x},\bar{z}). 
\end{equation}
Substituting this in eq. (\ref{3.21}) we obtain the following (exact)
equation for $\eta$
\begin{eqnarray}
\label{3.24}
&&\displaystyle\bar{\nabla}^2_{\mu}\eta +N^2(\eta)\eta - \frac{N^2(\eta)\beta}{2}\left[\mu^2(\eta_{\bar{x}})^2+ 2\mu^2G^{12}\eta_{\bar{x}}\eta_{\bar{z}} + \left(\frac{1}{G}+\mu^2(G^{12})^2 \right)[(\eta_{\bar{z}})^2 +2\eta_{\bar{z}}] \right] \\ \nonumber
&&=-\mu^2\left(\displaystyle\frac{\partial G^{12}}{\partial\bar{x}}
+ G^{12} \displaystyle\frac{\partial G^{12}}{\partial\bar{z}}\right)+
N^2(\eta)\left\{h(\bar{x})(1-\frac{\bar{z}}{H})+\frac{\beta}{2}\left[(\frac{1}{G}+\mu^2(G^{12})^2)-1\right] \right\} . 
\end{eqnarray}

\setcounter{equation}{0}
\section{Analytic solutions of Long's equation}

In the traditional representation of Long's equation the topography 
determines the shape of the flow domain and as a result it is not 
feasible to obtain analytic solutions to this equation even in some 
limits of the parameters $\beta$ and $\mu$ . We now show that this problem 
can be overcome in some limiting cases when the terrain following formulation
of this equation is used.

We consider two limiting cases $\beta=0,\mu=0$ and $\beta \ne 0,\mu=0$
we also assume $N^2(\psi)= \mbox{constant}$. For brevity we drop in the 
following the bars over $x,z$.

\subsection{The Limiting Case $\beta=0,\mu=0$}

In this case eq. (\ref{3.21}) simplifies to 
\begin{equation}
\label{4.1}
\frac{\partial^2\psi}{\partial z^2} + GN^2\psi = GN^2\left[ z +
h(x)(1-\frac{z}{H})  \right] 
\end{equation}
whose general solution is
\begin{equation}
\label{4.2}
\psi= A(x)\cos(\nu z)+ B(x)sin(\nu z)+ \left[z + h(x)(1-\frac{z}{H})\right] .
\end{equation}
Here $\nu =N\sqrt{G}$ and $A(x),B(x)$ are functions which have to be
determined from the boundary conditions . 

The boundary condition (\ref{3.22}) implies $A(x)=-h(x)$. To determine 
$B(x)$ we must apply the radiation boundary condition as 
$z \rightarrow \infty$ on the solution. To this end we must insure that
the vertical group velocity of the wave is positive. Using the dispersion
relation for hydrostatic flow given in [16,p.181] this group velocity
is: 
\begin{equation}
\label{4.3}
c_g=\frac{Nk\, sgn (\nu)}{\nu^2}
\end{equation}
where $k$ is the horizontal wave number. We deduce then that the 
vertical group velocity is positive when $k\nu \ge 0$.

To impose this condition on the solution (\ref{4.2}) we express $A(x),B(x)$
in Fourier integral form
\begin{equation}
\label{4.4}
A(x)= \int_{-\infty}^{\infty} a(k)e^{ikx} dk,\,\,\,B(x)=
\int_{-\infty}^{\infty} b(k)e^{ikx} dk
\end{equation}
where $k$ is the horizontal wave number. We deduce then that the 
solution  (\ref{4.2}) can be written as
\begin{equation}
\label{4.5}
\psi=\frac{1}{2}\left\{\int_{-\infty}^{\infty}(a(k)-ib(k))e^{i(kx+\nu z)} dk +\int_{-\infty}^{\infty}(a(k)+ib(k))e^{i(kx-\nu z)} dk\right\} + 
\left[z + h(x)(1-\frac{z}{H})\right]
\end{equation}
To satisfy the radiation boundary condition for $z \rightarrow \infty$ 
the first and second integral must vanish for $k < 0$
and $k > 0$ respectively. Therefore $a(k)$ and $b(k)$ must satisfy
\begin{equation}
\label{4.6}
a(k)=-i\,sgn (k) b(k)
\end{equation}
which implies that $B(x)$ is the Hilbert transform of $A(x)=-h(x)$. This
represents a complete analytic solution of Long's equation for this 
limiting case. 

We compare now these analytic results with the solution methodology that
has been used previously in the literature as was discussed  $Sec\; 2$.
First we note that this analytic solution requires only the direct (and simple)
computation of the Hilbert transform of the terrain function $h(x)$ .
This is a straightforward procedure even if it has to be done numerically.
On the other hand to compute $q(x)$ using (\ref{2.34}) requires
in general the solution of an integral equation. To do so one must use an
iterative algorithm which might turn out to be unstable or non-convergent
over complex terrain. Furthermore there is the issue
of applying the boundary conditions on $\psi$ at the terrain. To this
end the procedure discussed in $Sec\; 2$ requires the use
of the approximations that lead to (\ref{2.31}). As a result
the equation that is used to compute $q(x)$ (eq. (\ref{2.34})) is
also an approximate equation which will yield at best approximate
solution for this function. On the other hand the application of the 
boundary conditions using the procedure discussed in this section is exact
and does not place constraints on the height of the terrain.

From an experimental geophysical point of view it has been a common practice
to assume that the gravity wave generated by a flow over terrain is of the
form $\sin(kx+mz)$ (or similar)[14,15,28]. This has led to difficulties in the
eduction of this wave from experimental data. Our results show that
this form of the wave is incorrect (at least in principle). Furthermore
the numerical simulations that we carry in the next section demonstrate
that complex terrain can alter drastically the shape and amplitude of this
wave due to interference effects.

\subsection{The Limiting Case $\beta \ne 0,\mu=0$}

In this limiting case eq. (\ref{3.21}) becomes
\begin{equation}
\label{4.7}
\frac{\partial^2\psi}{\partial z^2} - GN^2 \left\{ -\psi +\frac{\beta}{2}
\left[\frac{1}{G}(\frac{\partial \psi}{\partial z})^2-1 \right] + 
z + h(x)(1-\frac{z}{H}) 
\right\} = 0. 
\end{equation}

Since this is a nonlinear equation we can find an approximate analytical 
solution using first order perturbation expansion under the assumption that
$\alpha^2= \frac{N^2\beta}{2} \ll 1$ (which is satisfied in most practical
situations). Expressing $\psi$ approximately as
$$
\psi =\psi_0 + \alpha^2 \psi_1
$$
and substituting this expression in (\ref{4.7}) we obtain 
to order zero and one in the parameter $\alpha^2$ the following equations  
\begin{equation}
\label{4.8}
\frac{\partial^2\psi_0}{\partial z^2} + GN^2\psi_0 = GN^2\left[ z +
h(x)(1-\frac{z}{H}) \right] 
\end{equation}
\begin{equation}
\label{4.9}
\frac{\partial^2\psi_1}{\partial z^2} + GN^2 \psi_1 -\alpha^2\displaystyle
(\frac{\partial \psi_0}{\partial z})^2 = -G\alpha^2 
\end{equation}

The boundary conditions on $\psi_0 , \psi_1$ are given by eq. (\ref{3.22}) at 
$z=0$ and radiation boundary conditions as $z \rightarrow \infty$.

Solving these (linear) equations for $\psi_0$ and $\psi_1$
we obtain the following expressions for their solutions 
\begin{equation}
\label{4.10}
\psi_0= A(x)\cos(\nu z)+ B(x)\sin(\nu z) + [z + h(x)(1-\frac{z}{H})] 
\end{equation}
\begin{equation}
\label{4.10a}
\psi_1= C_1(x)\cos(\nu z)+ C_2(x)\sin(\nu z) + f_1(x,z)+f_2(x,z)+f_3(x) 
- \frac{\beta}{2} 
\end{equation}
where
$$
f_1(x,z)=\frac{(A^2(x)-B^2(x))cos(2\nu z)+ 2A(x)B(x)sin(2\nu z)}{6} 
$$
$$
f_2(x,z)=\frac{(H - h(x))((\nu A(x)z+B(x))\cos(\nu z)+\nu B(x)z\sin(\nu z)}{H\nu} 
$$
$$
f_3(x)=\frac{A(x)^2+B(x)^2}{2}+\frac{1}{\nu^2}\left[1-\frac{h(x)}
{H}\right]^2.
$$
The determination of the functions $A(x),B(x),C_1(x),C_2(x)$ from the boundary 
conditions can be done using the same procedure outlined in the previous 
subsection.  However as it is algebraically cumbersome we omit 
the details.

\setcounter{equation}{0}
\section{Numerical Simulations over Complex Terrain.}

In the previous section we presented analytical solutions to 
Long's equation in some limiting cases. In general  
one has to resort to numerical simulations of Long's equation.
However as it was said in the introduction the terrain following form
of Long's equation obviate the need to discretize the terrain 
in these numerical procedures and therefore lead to better 
representations of complex terrains. Also the application of the
boundary condition at the terrain is being simplified considerably
and is treated exactly contrary to the approach discussed in $Sec\;2$
(see eqs. (\ref{2.31}) and (\ref{2.34})).

These simulations are carried in order to validate our equations and 
explore the different flow patterns (and possible special effects) that 
can be predicted over complex topography. 
The geophysical motivation for these simulations is related to recent 
experiments to measure and educe gravity waves and their properties
over the Alps from balloon data. This endeavor faced several 
difficulties. In part these difficulties can be traced to the fact that 
the terrain over which the measurements were made contained (at least) two 
summits rather than one [15].

In this section we carry simulations over three terrains. These consist 
of one, two and three hills with the following shape functions
\begin{equation}
\label{5.1}
f(x) = \frac{1}{(1+x^2)^{3/2}} ,
\end{equation}
\begin{equation}
\label{5.2}
f(x) = \frac{1}{(1+x^2)^{3/2}} +\frac{1}{(1+(x-5)^2)^{3/2}} ,
\end{equation}
\begin{equation}
\label{5.3}
f(x) = \frac{1}{(1+x^2)^{3/2}} + \frac{1}{(1+(x-5)^2)^{3/2}} + \frac{1}{(1+(x-10)^2)^{3/2}}.
\end{equation}
In all cases we solved (\ref{3.24}) for the perturbation
with the following parameters; 
\begin{equation}
\label{5.4}
\tau = 0.25 ,\;\; N=1,\;\;\beta = 1.10^{-2},\;\;\; \mu=0.1  .
\end{equation}
Here $N$ represents the nondimensional Brunt-V\"{a}is\"{a}l\"{a} frequency 
which was defined in (\ref{2.22}).

To solve for the perturbation $\eta$ over a finite two dimensional 
domain which after the transformation (\ref{3.1}) is represented by
$[a,c] \times [0,H]$ we imposed at the boundary $z=0$ 
the condition $\eta=0$. Radiation boundary conditions were imposed at $x=c$ 
and $z=H$. These are necessary to avoid reflection of the outgoing wave. 
To implement these boundary conditions we used "sponge boundaries" at 
$x=c$ and $z=H$ [23,24]. The sponge damping constant was calibrated to 
suppress wave reflection at these boundaries. In addition open boundary 
conditions were used at these boundaries themselves viz. we let
$$
\frac{\partial \eta}{\partial x}(c,z) =0,\,\,\,
\frac{\partial \eta}{\partial z}(x,H) =0.
$$

In the simulations described 
below we let (in nondimensional units) $a=-25$, $c=25$ and $H=25$ with a
grid of $801 \times 401$ points. The sponge layer (at $a,c$) consists of 
$20$ grid points.  

To solve (\ref{3.24})  under these settings we  
used Matlab[28]. Central finite differences approximations were used 
to discretize Long's equation on the domain grid and a fixed point 
iterative algorithm was implemented to solve the resulting equations. 
The convergence criteria for the iterations 
was that the step error $|\eta_{m+1} -\eta_{m}|$ was less 
than $1.10^{-9}$ where $m$ is the iteration number. Convergence
was achieved in less than $100$ iterations. 
----------------------------------------------------------
It should be noted however that $\eta$ in these
simulations represents the perturbation from the "terrain following stream
function" which was defined in (\ref{3.8}). Using (\ref{3.6}),
(\ref{3.8}) we find that the perturbation to the
base flow field $u_p, w_p$ is given by
\begin{equation}
\label{5.5}
u_p(x,z)= \frac{h(x)}{H-h(x)}+\frac{H}{H-h(x)}\frac{\partial\eta}
{\partial{\bar{z}}}(x,\bar{z})
\end{equation}
\begin{equation}
\label{5.6}
w_p(x,z)= -\frac{\partial\eta}{\partial{\bar{x}}}(x,\bar{z}) -
G^{12}(x,\bar{z})
\left(1+\frac{\partial\eta}{\partial{\bar{z}}}(x,\bar{z})\right)
\end{equation}
where $\bar{z}$ is defined in (\ref{3.1}).
From this vector field we can  compute the "regular" stream function
$\phi(x,z)$ which is defined as
\begin{equation}
\label{5.7}
\phi(x,z)=z+\nu(x,z)
\end{equation}
where $\nu(x,z)$ satisfies
\begin{equation}
\label{5.8}
u_p=\frac{\partial\nu}{\partial z},\,\,\, w_p=-\frac{\partial\nu}{\partial x}
\end{equation}

A zoom-plot of $\phi(x,z)$ (on part of the computational domain)
on the region around the three topographies discussed above is presented
in $Figs\,1,2,3$.  We observe that the phase lines of the stream function
in all these figures tilts with height. In all three figures the gravity
waves are in the lee of the topography.

These figures demonstrate the impact that complex terrain can have on
the structure of gravity waves.

$Fig\;1$ displays the results of the simulation for one hill (described
by (\ref{5.1})). The plot corresponds to the classical results that appeared
in the literature for this case. The results for two hills in $Fig\;2$
show some wave activity between the two hills and less organized waves
with somewhat smaller amplitude in the lee of the two hills. This might be
due to interference effects between the waves generated by the two hills.
For three hills the simulation shows the same wave activity between the
two hills (on the left) as in $Fig\;2$ but then a "quiet" zone between the
second and third hills in which $\phi$ is almost constant. The waves
in the lee of the three hill appear regular as in $Fig\;1$.

\setcounter{equation}{0}
\section{Summary and Conclusions.}

We derived in this paper a terrain following formulation of Long's equation
in which the topography is "absorbed" in the coefficients of
the differential equation representing the flow rather than being
part of the boundary conditions. We used this representation
to solve Long's equation analytically in some limiting cases and 
numerically over complex topography. The new formulation also opens  
the possibility to develop analytical estimates which compare the 
solutions of this equation over different topographies .

From a geophysical point of view it well known that some present models 
for the generation of gravity waves over estimate this effect [25,27]. 
Partially, this is due to the fact that they use oversimplified 
representation of the terrain. Furthermore they do not take into 
account the effects that are due to complex terrain (as demonstrated 
by our simulations). We believe that the new form of Long's equation will 
make it easier to consider more realistic representations of the terrain and 
its effect on the generation and propagation of gravity waves.

\newpage
\centerline{\Large{\bf List of Captions}}

\vspace*{.50in}

\begin{tabular}{ll}

Fig. 1  Zoom-Contour plot of the stream function $\phi$ over one hill \\
centered at $x=0$ with $N=1$, $\beta=0.01$, $\mu=0.3$, $\tau=0.25$. \\
\\
Fig. 2  Same as $Fig\, 1$  with two hills centered at $x=0$ and $x=5$. \\
\\
Fig. 3  Same as $Fig\, 1$ with three hills centered at $x=-5,0$ and $5$.

\\
\end{tabular}

\newpage
\includegraphics[scale=1,height=160mm,angle=0,width=180mm]{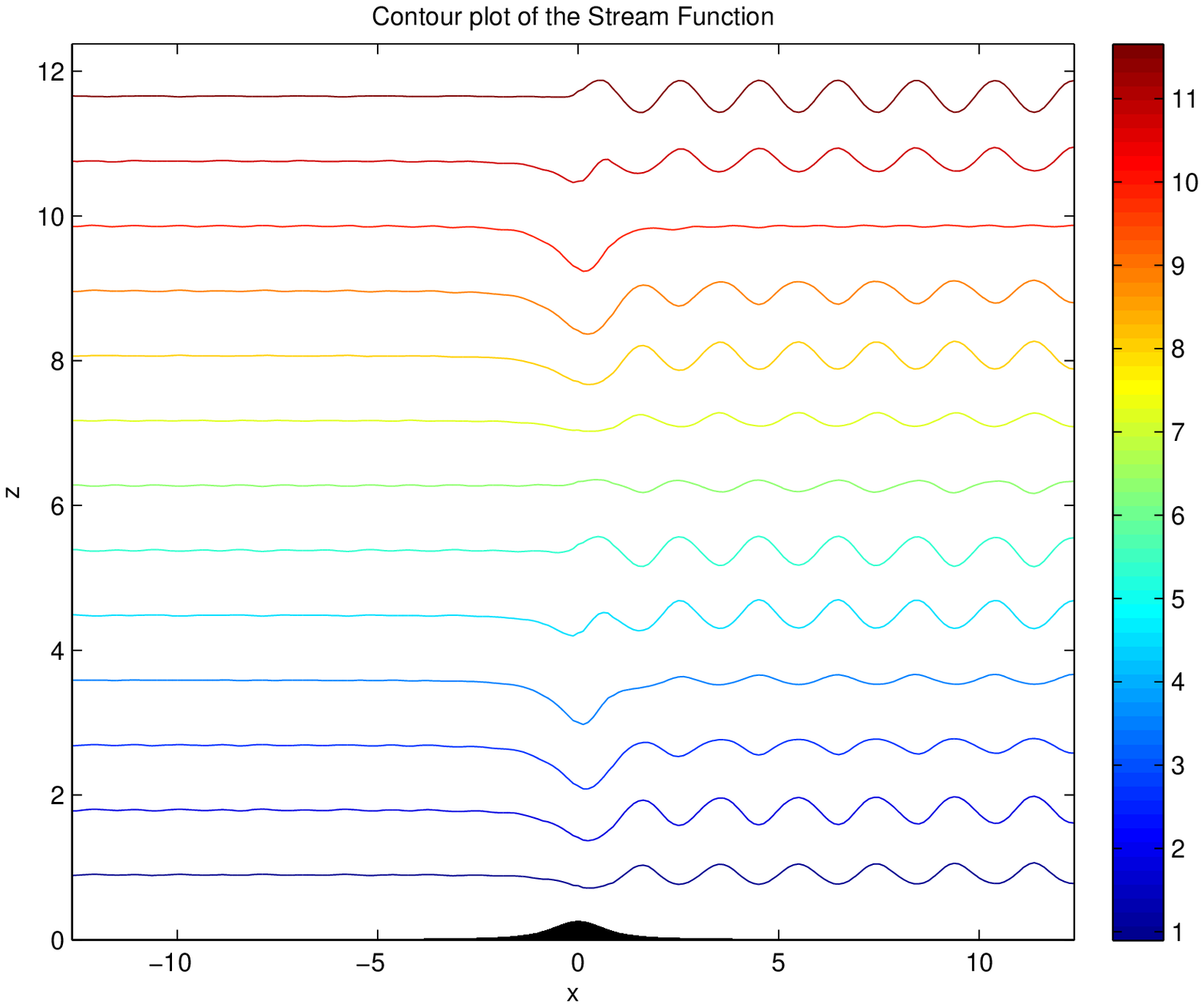}

\newpage
\includegraphics[scale=1,height=160mm,angle=0,width=180mm]{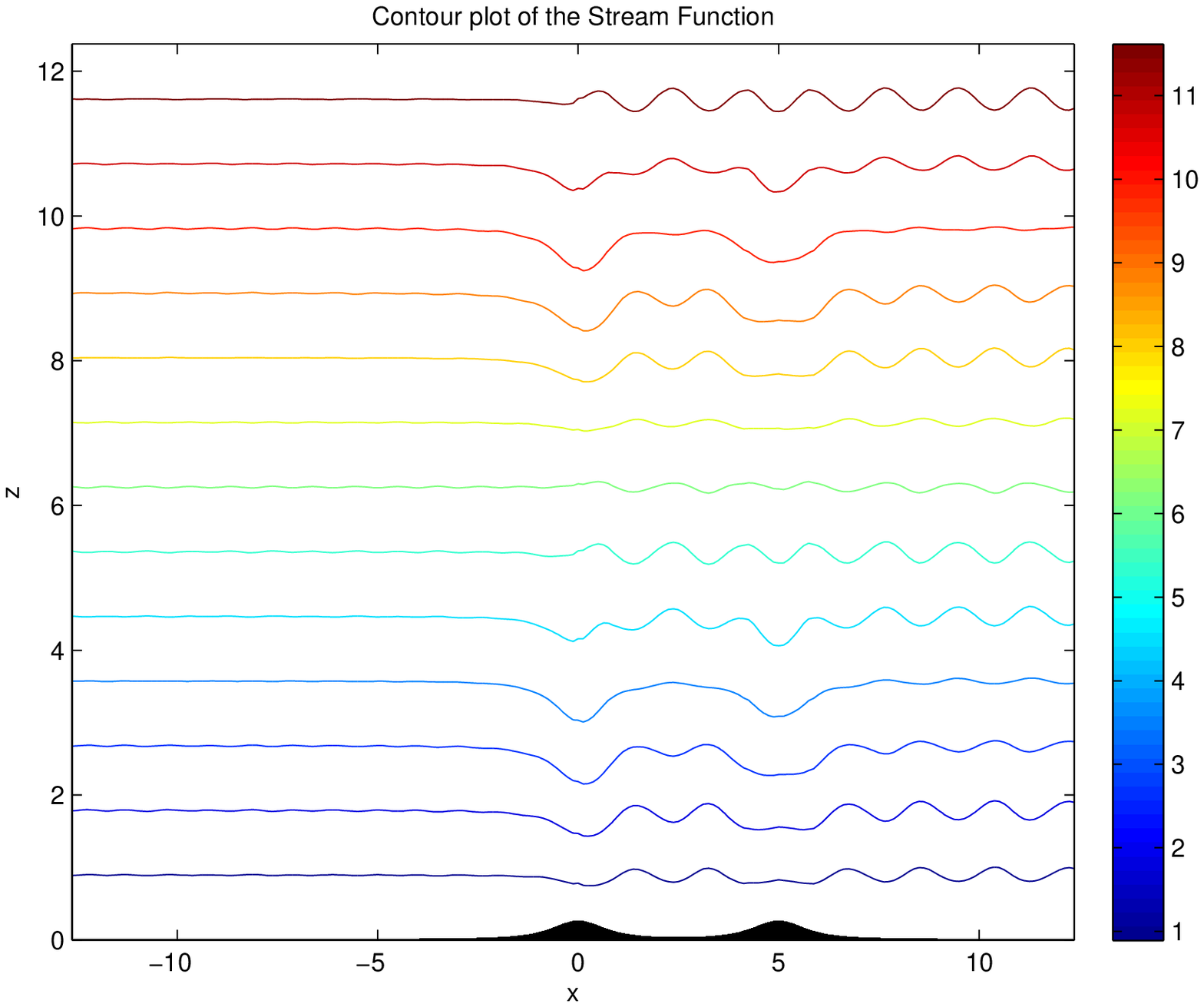}

\newpage
\includegraphics[scale=1,height=160mm,angle=0,width=180mm]{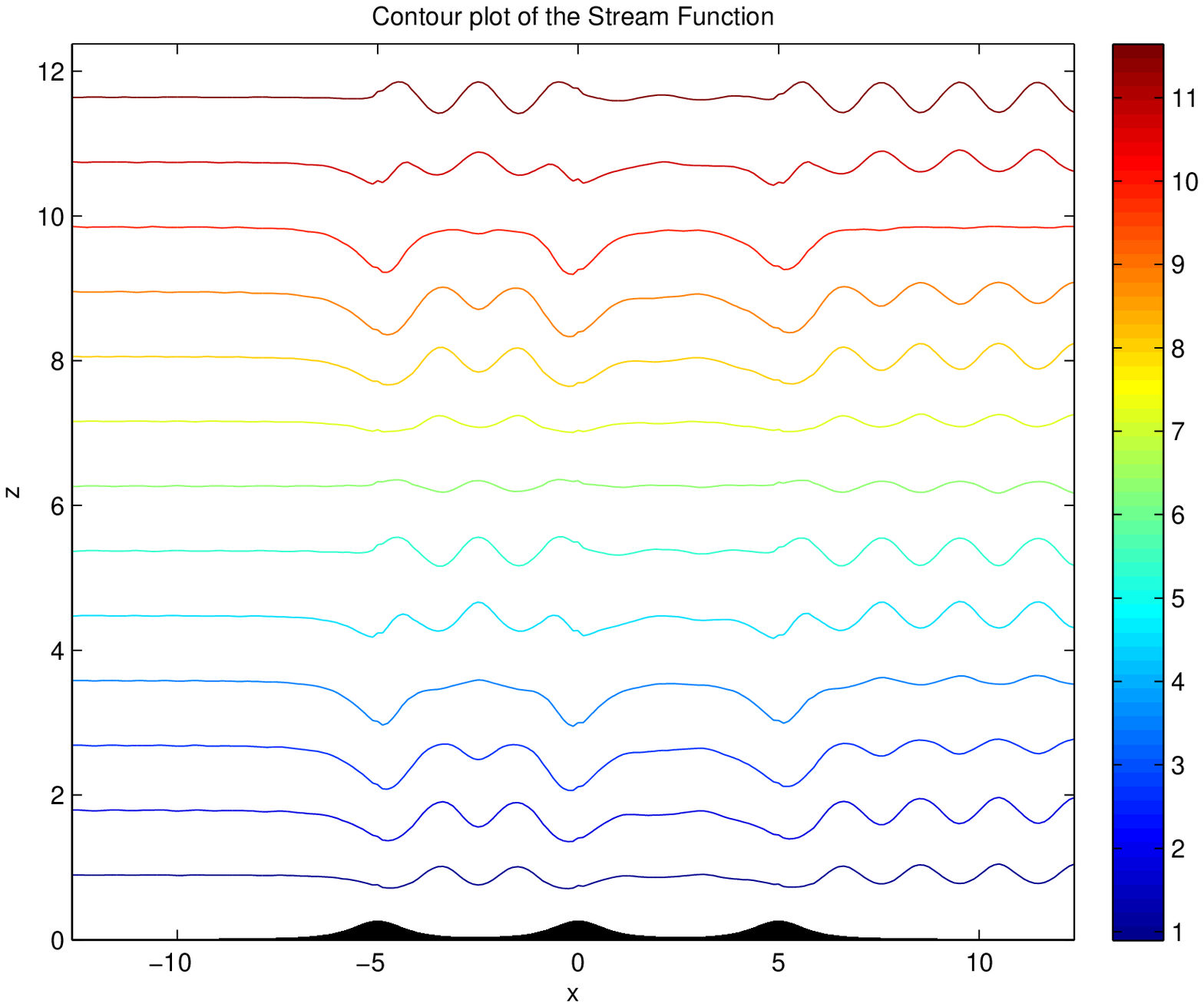}

\end{document}